\newcommand{\ex}{\mathrm{e}}
\newcommand{\dd}{\mathrm{d}}
\newcommand{\GN}{G_{_\mathrm{N}}}

\newcommand{\Ka}{\mathcal{K}}
\newcommand{\Hu}{\mathcal{H}}

\documentclass[universe,conference report,accept,oneauthor,pdlaftex,10pt,a4paper]{mdpi} 

\firstpage{1} 
\pubvolume{xx}
\issuenum{1}
\articlenumber{1}
\pubyear{2018}
\copyrightyear{2018}
\history{Received: 13 July 2018; Accepted: 09 August 2018; Published: date}
\updates{yes} 
\usepackage[utf8]{inputenc}
\usepackage{bm}

\Title{Using Trajectories in Quantum Cosmology}

\Author{Patrick Peter$^{1,2}$}

\AuthorNames{Patrick Peter}

\address{%
$^{1}$ \quad Institut d'Astrophysique de Paris (${\cal
	G}\mathbb{R}\varepsilon\mathbb{C}{\cal O}$), UMR 7095 CNRS; 98bis boulevard Arago, 75014 Paris, France.\\
$^{2}$ \quad Department of Applied Mathematics and Theoretical Physics, Centre for Mathematical Sciences,
\mbox{University of Cambridge}, Wilberforce Road, Cambridge CB3 0WA, UK}

\simplesumm{I discuss the application of quantum trajectory formalism to
the FLRW Universe.}
\abstract{
Quantum cosmology based on the Wheeler De Witt equation represents a
simple way to implement plausible quantum effects in a gravitational
setup. In~its minisuperspace version wherein one restricts attention to
FLRW metrics with a single scale factor and only a few degrees of freedom
describing matter, one can obtain exact solutions and thus acquire full
knowledge of the wave function. Although this is the usual way to treat a
quantum mechanical system, it turns out however to be essentially
meaningless in a cosmological framework. Turning to a trajectory approach
then provides an effective means of deriving physical consequences.
}


\begin{document}

\section{Introduction}

Quantum cosmology \cite{Halliwell:1990uy,Bojowald:2015iga} aims at
understanding how gravitational fields describing cosmological setups,
usually treated as purely classical backgrounds \cite{PeterUzan2009},
may be affected by quantization. In~particular, the most important issue
not addressed by classical general relativity, i.e., the singularity from
which our Universe ensues, could be tackled by imposing physically
relevant boundary conditions on the wave function.

The Universe being by definition unique, and quantum measurements being
understood by means of ensemble averages, i.e., repeated experiments, the
meaning of the wave function of the Universe seems rather unclear. There
exists however a formulation of quantum mechanics, originally developed
by de Broglie \cite{deBroglie:1925cca} and Bohm
\cite{Bohm:1951xw,Bohm:1951xx} and based on trajectories
\cite{Holland:1993ee} that, as it happens, is easily applicable to
cosmology \cite{Pinto-Neto:2013toa}. It is in this framework that one
can assign actual values at each instant of time to the scale factor
(the quantum trajectory) \cite{AcaciodeBarros:1997gy,Peter:2016kan} and
even address the question of time \cite{AcaciodeBarros:1998nb}.

In the following, I briefly recap how gravitation may be quantized {\sl
\`a la} Wheeler De Witt and how does the restriction to
Friedmann-Lema\^itre-Roberston-Walker (FLRW) minisuperspace
provides a time-dependent Schr\"odinger-like equation when a perfect
fluid is considered to be the source of Einstein equations. The trajectory
approach then permits to derive a fully quantum time-dependent scale
factor whose properties are examined in detail.

\section{General Setup}
\vspace{-6pt}

\subsection{Classical Hamiltonian General Relativity}

\begin{figure}[H]
\centering \includegraphics[width=7 cm]{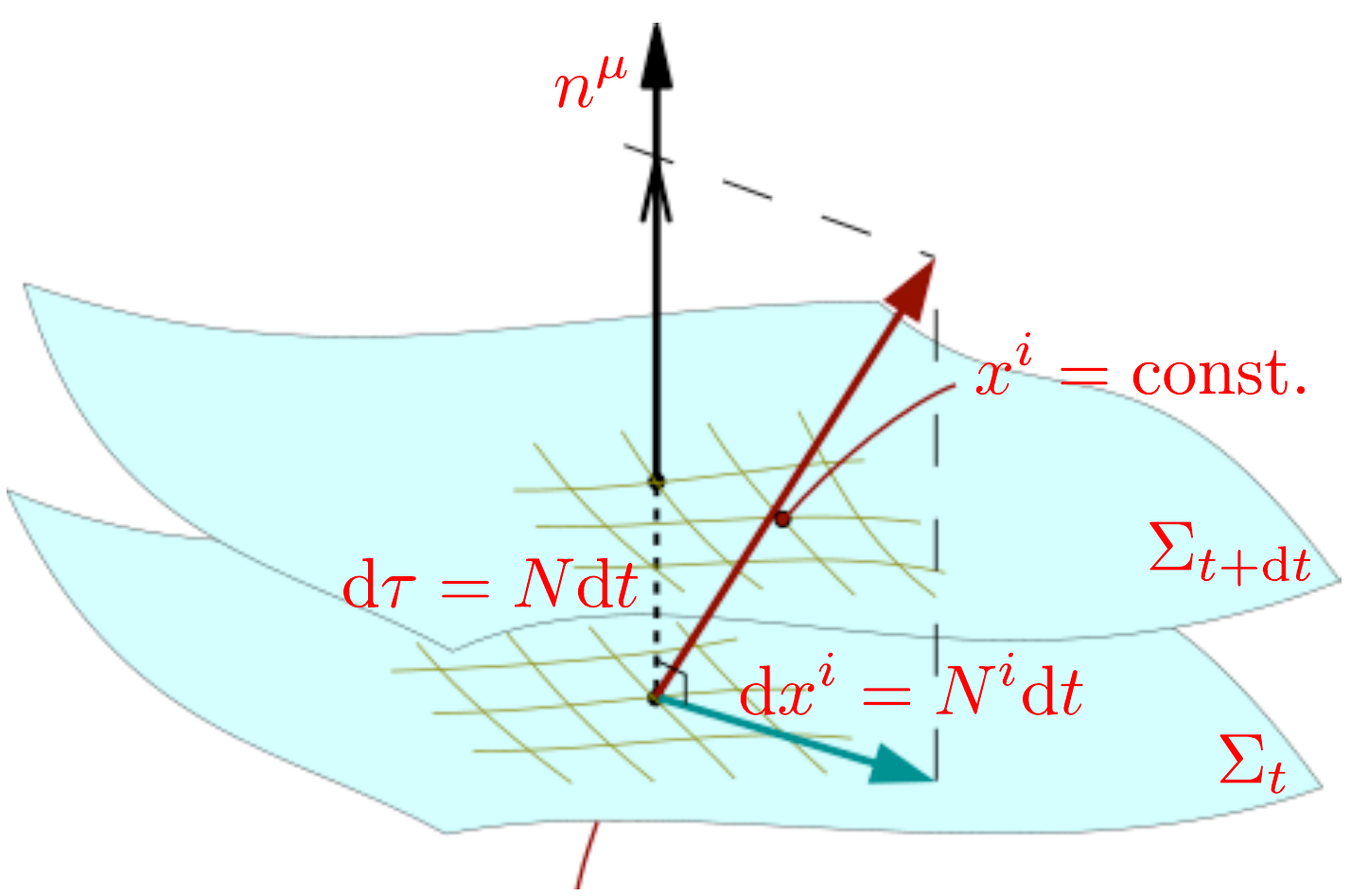} \includegraphics[width=7cm]{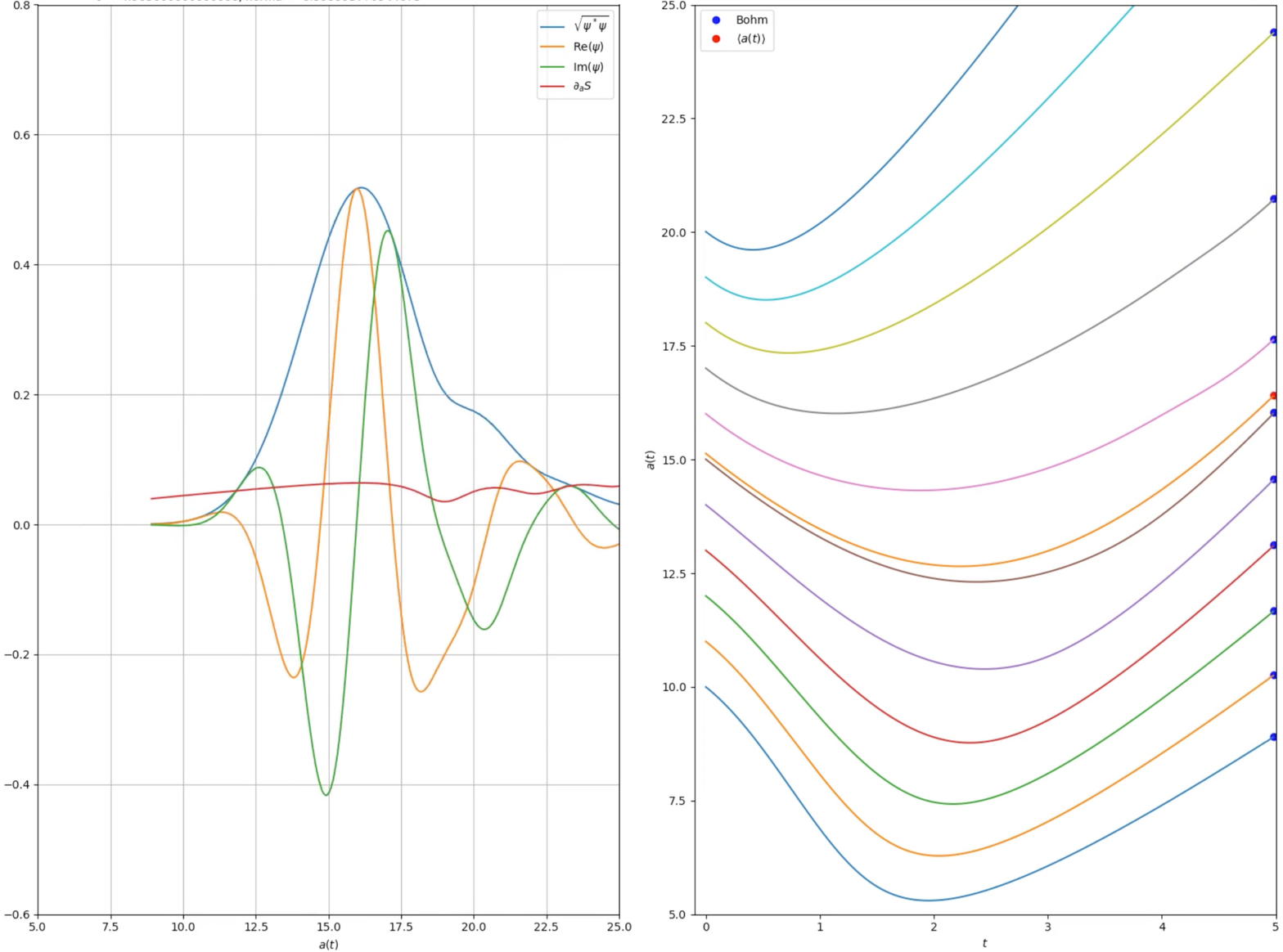}
\caption{ (\textbf{a}) Spacetime foliation in terms of hypersurfaces
	$\Sigma_t$, each labelled by a time-like parameter $t$. The diagram
	defines the normal $n^\mu$ and makes explicit the decomposition of a
	tangent to the worldline $x^i=\mathrm{const}$ through
	the lapse function $N$ and shift vector $N^i$.  (\textbf{b}) A
	time-evolved example with a few typical trajectories showing a quantum
	non-singular bouncing universe. }\label{ADM}
\end{figure}

Since the purpose is to quantize general relativity (GR)
\cite{Kiefer:2007ria}, one starts from the usual Einstein-Hilbert action
on a compact space $\mathcal{M}$ with boundary $\partial \mathcal{M}$,
including a possible cosmological constant $\Lambda$,
\begin{equation}
\mathcal{S} = \frac{1}{16\pi\GN} \left[ \int_\mathcal{M} \!\!\! \sqrt{-g}
\ \left( R -2\Lambda \right)\dd^4 x + 2 \int_{\partial \mathcal{M}} \!\!
\!\sqrt{h} K^i_{\ i} \dd^3 x \right] + \mathcal{S}_\mathrm{matter} \left[
\Phi\left(x\right)\right],
\label{EH}
\end{equation}
where the Ricci scalar $R$ is coupled to matter fields symbolically
named $\Phi$. Figure \ref{ADM} shows the usual 3+1 split of spacetime
when the metric takes the form
\begin{equation}
\dd s^2 = g_{\mu\nu} \dd x^\mu \dd x^\nu = - N^2 \dd t^2
+ h_{ij} \left( \dd x^i + N^i \dd t\right) \left( \dd x^j + N^j \dd t\right).
\label{Metric}
\end{equation}

In \eqref{EH}, the extrinsic curvature of each leaf $\Sigma_t$ is given
by
\begin{equation}
K_{ij} = - \nabla^{(h)}_j n_i = \frac{1}{2 N} \left( \nabla^{(h)}_j N_i +
\nabla^{(h)}_i N_j - \frac{\partial h_{ij}}{\partial t}\right),
\label{Kij}
\end{equation}
where $\nabla^{(h)}$ is the covariant derivative associated with
the intrinsic metric $h_{ij}$. From 
\begin{equation}
\mathcal{S} = \int L\dd t = \frac{1}{16\pi\GN} \int \dd t \,
\left[\int \dd^3 x\, N\sqrt{h} \left(
K_{ij} K^{ij} -K^2 +^3\!\! R-2\Lambda \right) + L_\mathrm{matter} \right],
\label{SLdt}
\end{equation}
one derives the canonical momenta
\begin{equation}
\pi^{ij} \equiv \frac{\delta L}{\delta \dot{h}_{ij}} = -\frac{\sqrt{h}}{16\pi\GN} \left( 
K^{ij} - h^{ij} K\right), \ \ \ \ \ \pi^{0} \equiv \frac{\delta L}{\delta \dot{N}} = 0, 
\ \ \ \ \ 
\pi^{i} \equiv \frac{\delta L}{\delta \dot{N^i}} = 0,
\label{Momenta}
\end{equation}
the last two providing primary constraints, as well as the momentum associated
with the matter component, $\pi^\Phi$ say. The Hamiltonian is therefore
\begin{equation}
H \equiv \int \dd^3 x \left( \pi^0 \dot{N} + \pi^i \dot{N}_i +\pi^{ij}
\dot{h}_{ij} + \pi^\Phi \dot\Phi \right) - L =  \int \dd^3 x \left(
\pi^0 \dot{N} + \pi^i \dot{N}_i + N H + N_i \Hu^i \right).
\label{Hamil}
\end{equation}

Variations of \eqref{Hamil} yields the Hamiltonian description of GR.

\subsection{Quantization}

Quantum mechanics proceeds by first defining a Hilbert space of
accessible states. In the GR case, it is the space of all the 3-metrics
$h_{ij}$ and matter fields compatible with diffeomorphism invariance; it~is called superspace. The wave functional is then $\Psi\left[
h_{ij}\left(x\right),\Phi\left(x\right)\right]$ and depends on the
coordinates $\{x^\mu\}$, now understood as mere parameters.

Upon adopting the Dirac canonical quantization procedure whereby
canonical momenta are replaced by $-i$ times the functional derivative
with respect to the variable they are the momenta of,~i.e.,
\begin{equation}
\pi^{ij} \to - i \frac{\delta}{\delta h_{ij}}, \ \ \ \ \ 
\pi^{0} \to - i \frac{\delta}{\delta N}, \ \ \ \ \ 
\pi^{i} \to - i \frac{\delta}{\delta N_i}, \ \ \ \ \
\pi^{\Phi} \to - i \frac{\delta}{\delta \Phi},
\label{Dirac}
\end{equation}
one finds that the primary constraints translate into the fact
that the wave function depends neither on the lapse function
nor on the shift vector, that it is unchanged under diffeomorphisms,
and finally that the Wheeler De Witt equation
\begin{equation}
H\Psi = \left[ -16\pi\GN \mathcal{G}_{ijkl}
\frac{\delta^2}{\delta h_{ij} \delta h_{kl}} + \frac{\sqrt{h}}{16\pi\GN}\left(
-^3R+2\Lambda + 16\pi\GN \hat{T}^{00}\right) \right]\Psi = 0
\label{WDW}
\end{equation}
holds, with the De Witt metric defined as
\begin{equation}
\mathcal{G}_{ijkl} = \frac{1}{2\sqrt{h}}\left( h_{ik} h_{jl} + h_{il} h_{jk}
-h_{ij} h_{kl} \right),
\label{DWm}
\end{equation}
and $^3R$ is the curvature associated with the metric $h_{ij}$. In
\eqref{WDW}, $\hat{\bm{T}}$ is the operator version on superspace of the
	stress energy tensor relevant for the matter fields.

\subsection{Minisuperspace}

As it is essentially out of question to solve \eqref{WDW} in general, one
restricts attention to the special FLRW case for which one replaces the
general 3D metric $h_{ij}$ by
\begin{equation}
h_{ij} \dd x^i \dd x^j \ \ \mapsto \ \ \ a^2(t) \left[  \frac{\dd r^2}{1-\Ka r^2} +
r^2 \left( \dd\theta^2 + \sin^2\theta \dd\phi^2 \right)\right],
\label{mini}
\end{equation}
leading to a numerical parameter, the spatial curvature $\Ka$, in
practice set to zero in agreement with observational data, and a
dynamical function, the scale factor $a(t)$. Under the assumption that
the 3D metric takes the form \eqref{mini}, the Wheeler De Witt equation
becomes a Schr\"odinger-like equation for, say, the 2 degrees of freedom
wave function $\Psi\left[ a(t),\phi(t) \right]$. There are many points,
both mathematical and physical, that can be raised about the
minisuperspace approach, but they shall not concern us in the framework
of this paper, and we refer the reader to, for instance,
Refs.~\cite{Kiefer:2008deh,Pinto-Neto:2013toa} and the references therein
for that matter.

We want instead here focus on the simplest possibility, namely that of
vanishing spatial curvature $\Ka\to 0$ and consider as the matter
component a perfect fluid which we treat using Schutz formalism~\mbox{\cite{Schutz1970,Schutz1971}}. In this formalism, the full Hamiltonian
reads
\begin{equation}
\dd s^2 = N^2(\tau) \dd\tau - a^2(\tau) \gamma_{ij} \dd x^i \dd x^j \ \ \ 
\hbox{and} \ \ \ p=w\rho \ \ \ \Longrightarrow \ \ \ 
H = N \left( - \frac{\pi_a^2}{4 a} + \frac{\pi_t}{a^{3w}}\right),
\label{Hfluid}
\end{equation}
where the variable $t$ is associated to the velocity potentials and we
keep the lapse function $N(\tau)$ unfixed for later convenience. With
the choice $N\to a^{3w}$ and for a radiation fluid having $w=\frac13$,
the~replacement $\pi_t\to -i\partial_t$ transforms the Wheeler De Witt
equation $H\Psi = 0$ into $i\partial_t\Psi = \frac14 \partial_a^2 \Psi$,
which~is, up to a sign, a time-dependent Schr\"odinger equation for a
free particle \cite{AcaciodeBarros:1997gy}. Not surprisingly, the~fluid
has permitted to define a global time variable.

\section{Quantum Trajectories}

Although it is not the purpose of this paper to review the trajectory method in
quantum mechanics, let me summarize it shortly.

Since we have seen that the matter content merely serves in our case to
define a time variable in the time-dependent Schr\"odinger equation, I
now consider a canonical transformation for the scale factor, namely
$\left(a ,\pi_a \right) \mapsto \left(A ,\Pi_A \right)$, leading to
$H\left(a,\pi_a\right)\mapsto \Hu \left(A,\Pi_A\right)$ by means of a
generating function $F(a,A;t)$ satisfying
\begin{equation}
\dot a \pi_a - H = \dot A \Pi_A - \Hu +\frac{\dd F}{\dd t} \ \ \ \ 
\Longrightarrow \ \ \ \ \dd F = \pi_a \dd a - \Pi_A\dd A +
\left( \Hu-H\right)\dd t.
\label{dFdt}
\end{equation}

From \eqref{dFdt}, one infers that 
\begin{equation}
\dd \left( F+A\Pi_A \right) = \pi_a \dd a + A\dd \Pi_A + \left( \Hu-H\right)\dd t,
\label{dS}
\end{equation}
showing that the function $\left( F+A\Pi_A \right)$ depends on $a$,
$\Pi_A$ and $t$.

We now choose the canonical transformation such that the new Hamiltonian $\Hu$
identically vanishes on shell. Hamilton equations then imply 
$$
\dot A = \frac{\partial \Hu}{\partial \Pi_A} = 0 \ \ \ \ \hbox{and} \ \ \ \
\dot \Pi_A = - \frac{\partial \Hu}{\partial A} = 0 \ \ \ \ \Longrightarrow
\ \ \ \ \frac{\dd \left(F+A\Pi_A\right)}{\dd t} = \dot a \pi_a - H = L,
$$
$L$ being the original Lagrangian [see Equation~\eqref{SLdt}]. Therefore, one
may identify the function $\left( F+A\Pi_A \right)$ with the action
$\mathcal{S} = \int L\dd t$. Equation~\eqref{dS}, taken on shell, now reads
\begin{equation}
\dd \mathcal{S} = \pi_a \dd a - H \dd t \ \ \ \ \Longrightarrow \ \ \ \
\frac{\partial \mathcal{S}}{\partial t} = - H \ \ \ \ \hbox{and} \ \ \
\ \frac{\partial \mathcal{S}}{\partial a} = \pi_a, \label{PW}
\end{equation}
which can be recast in the more usual Hamilton-Jacobi form
\begin{equation}
H\left( a,\frac{\partial \mathcal{S}}{\partial a} \right) +
\frac{\partial S}{\partial t} = 0, \label{HJ}
\end{equation}
and the last equation of \eqref{PW} relates the actual value of the
momentum to the gradient of the action; as~we shall see below, this will
be equivalent to the pilot-wave equation when the action is identified
with the phase of the wave function.

In a quantum framework, the modified Hamilton-Jacobi equation is
obtained as the real part of the Schr\"odinger equation when the wave
function is written explicitly as an amplitude $R\in \mathbb{R}$ and a
phase $S\in \mathbb{R}$, namely when setting $\Psi(a,t) = R \ex^{i S}$.
This is one way to identify the phase of the wave function with the
action. As a result, it is natural to assume that an actual
trajectory can be obtained as the solution of the canonical
transformation eikonal relation $\pi_a|_\mathrm{actual} \propto \dot
a = \partial S/\partial a$, the dot denoting a time derivative.

It is instructive to note as well that setting $\rho = R^2$ and
replacing the time derivative $\dot a$ by an actual velocity $v$, one
finds that the imaginary part of Schr\"odinger equation may be written
as
$$
\frac{\partial \rho}{\partial t} + \mathrm{div} \left( \rho v\right) =0,
$$
where the divergence is with respect to the variable $a$. In other
words, this formulation of quantum physics is very close to ordinary
fluid mechanics, and the same methods (Eulerian or Lagrangian) actually
apply.

Solving the trajectory equation is all but trivial, and many methods,
mostly numerical, have~been devised \cite{Wyatt:2005uc}. The one
we use in the examples presented below is based on a simple radial
basis interpolation, but it can be extended to include a moving mesh
method. In fact, if the initial distribution of points $\{a_n|_{t=t_0}\}$
at which the trajectories are calculated is $|\Psi|^2$, then it remains
distributed along the square of the wave function at all subsequent
times, so that whatever the behavior of $\Psi$, one is sure to
cover a domain that always remains where the wave function is
large. For our illustrative purpose however, this refinement is
not necessary.

\begin{figure}[H]
\centering
\includegraphics[width=6cm]{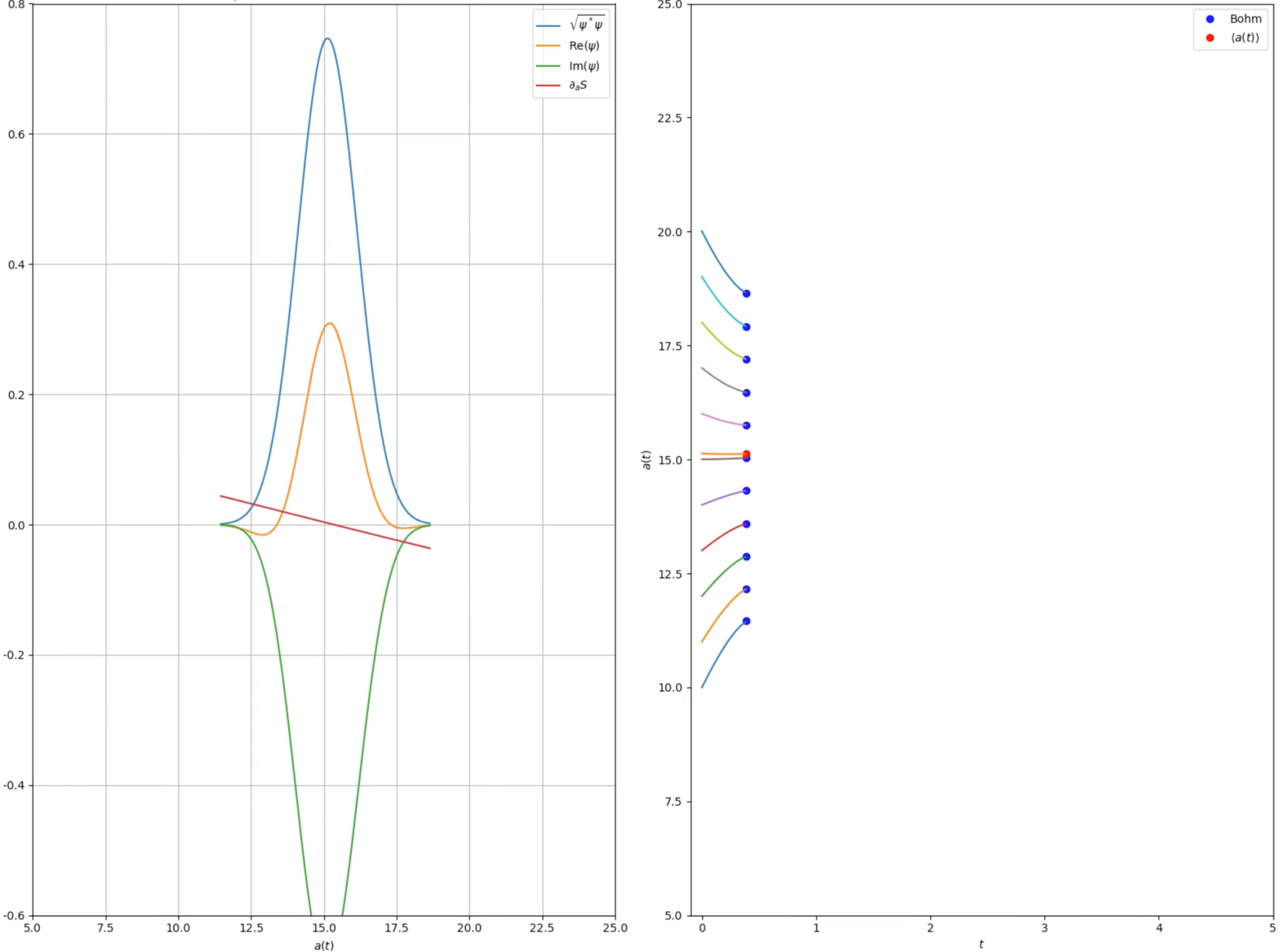}
\includegraphics[width=6cm]{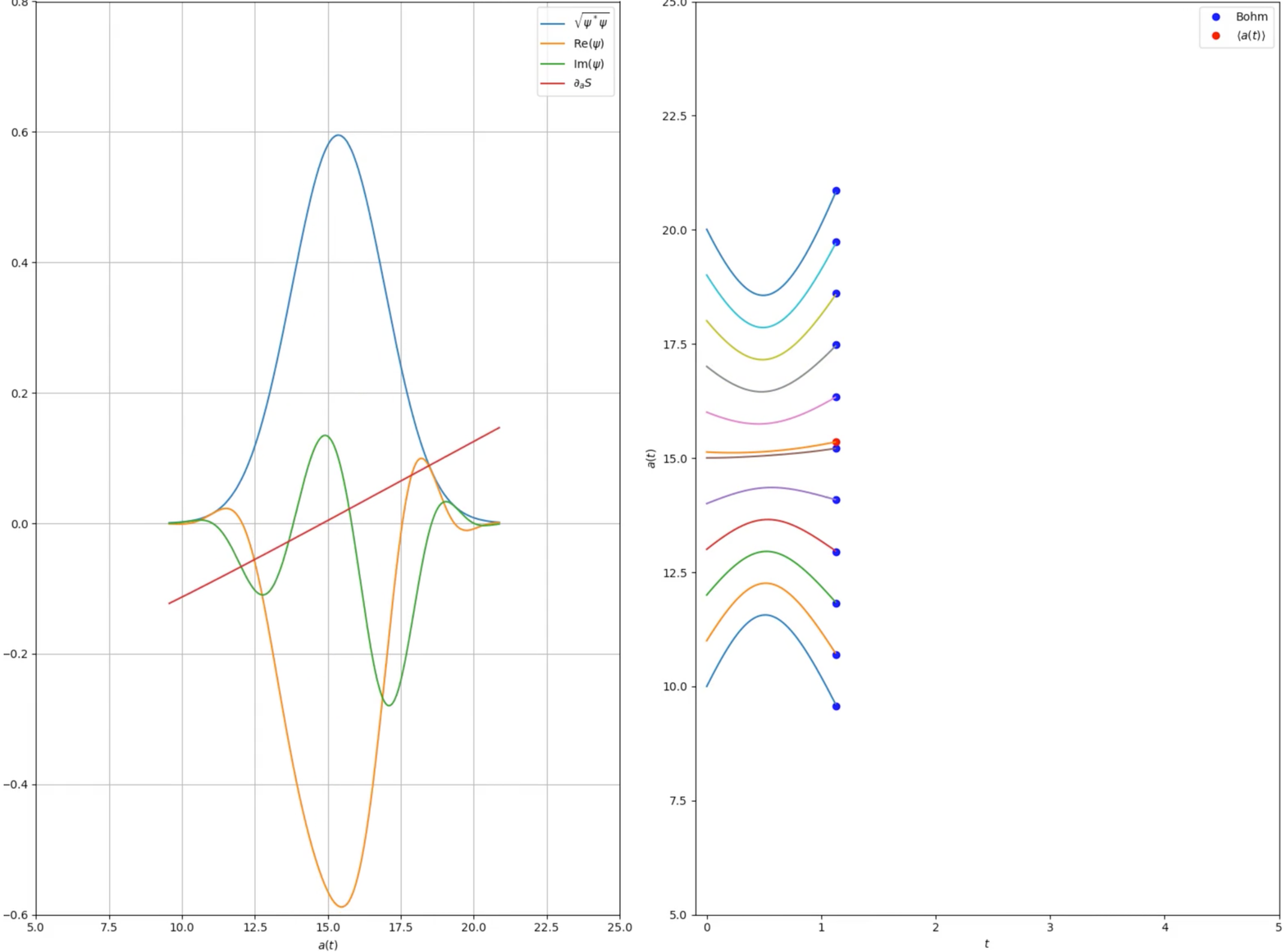}
\includegraphics[width=6cm]{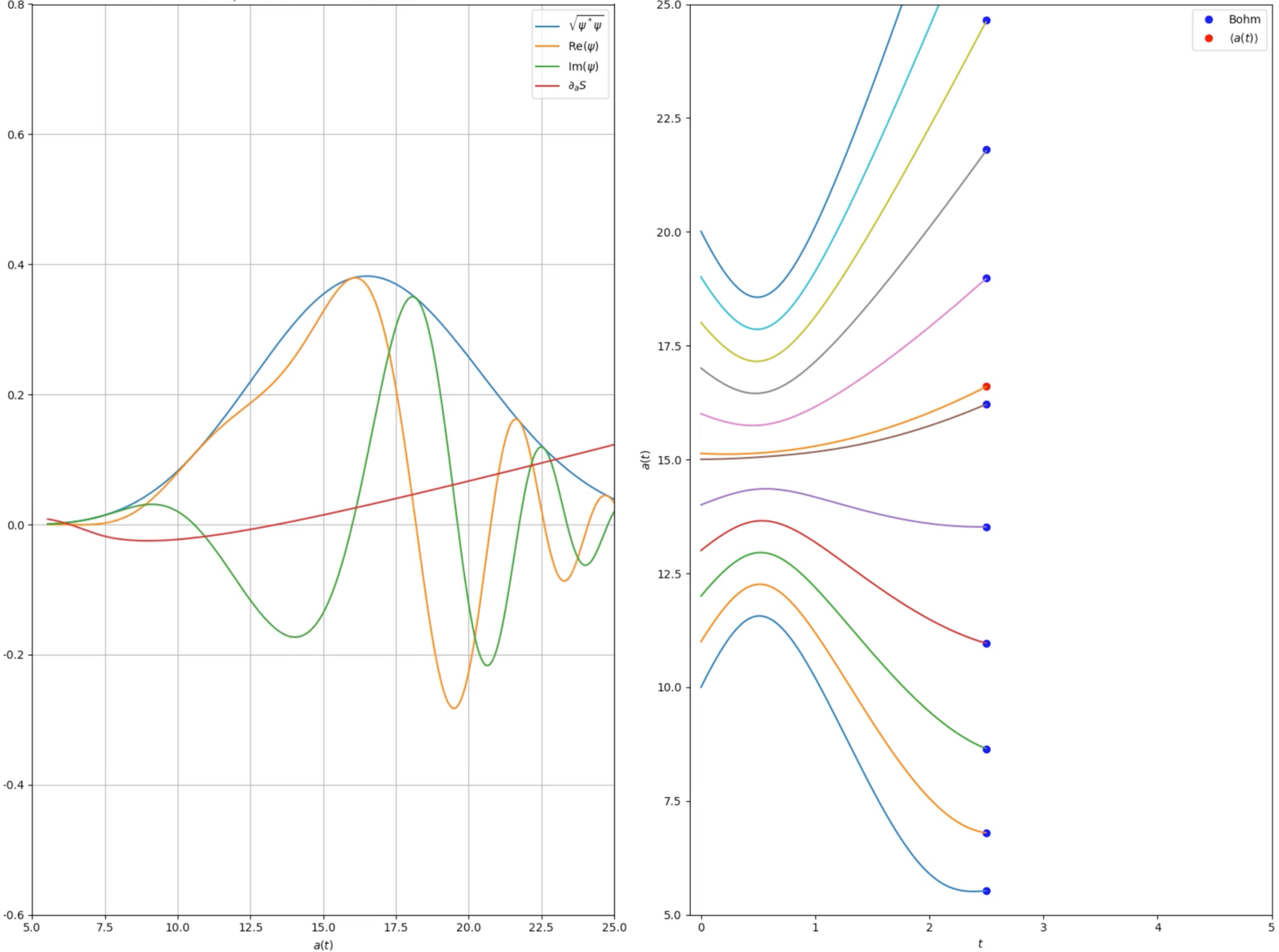}
\includegraphics[width=6cm]{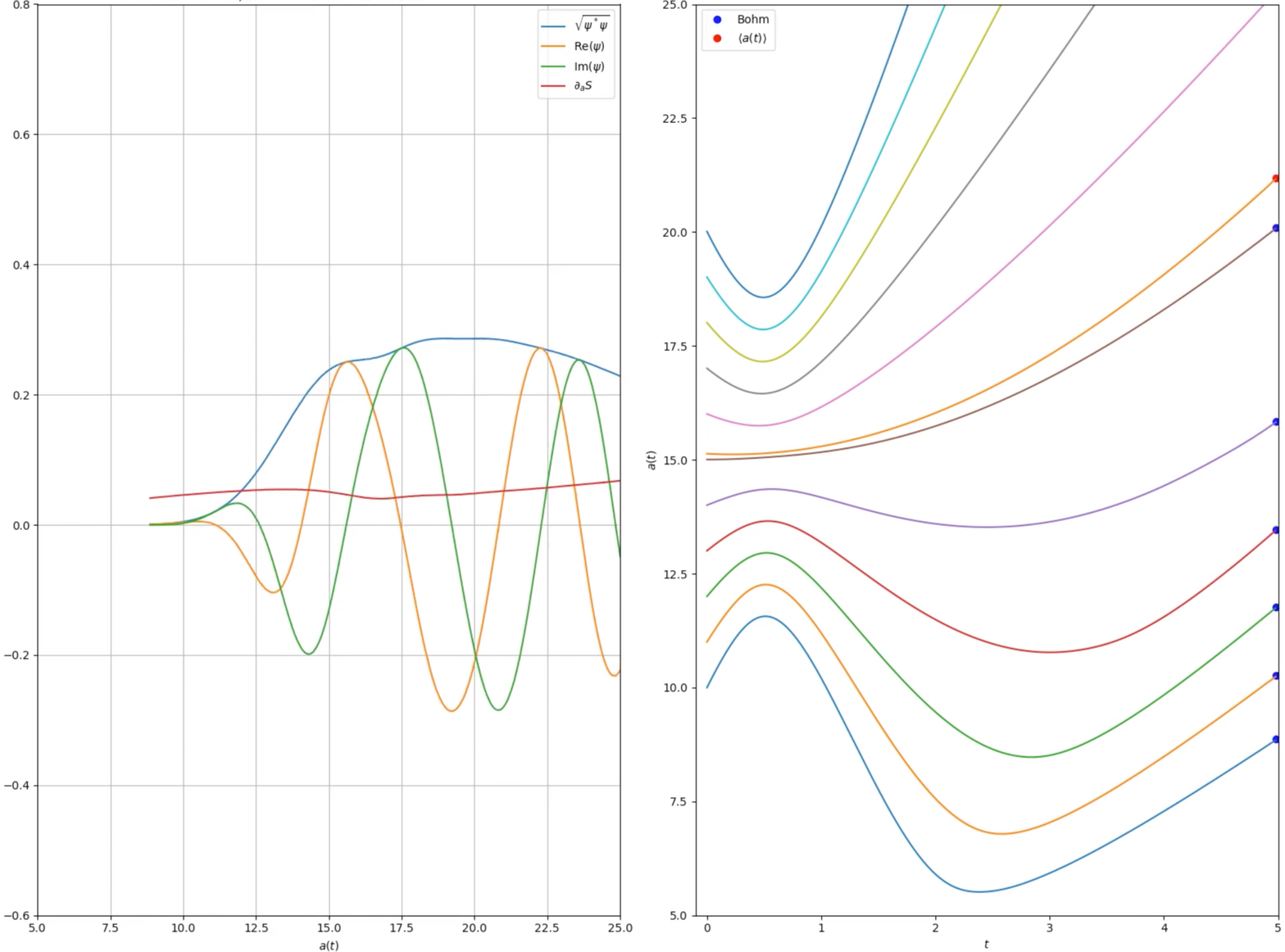}
\caption{Time evolution of a wave function and typical trajectories:
(\textbf{a}) initial condition, the phase gradient has both
positive and negative values, so some trajectories are expanding while
other are contracting, depending on their initial value. (\textbf{b}) An
instant later, the phase gradient has essentially reversed, so all
trajectories have bounced in one way or another. (\textbf{c}) After some
time, the~phase gradient tends to increase to become almost everywhere
positive. (\textbf{d}) Finally, the Universe starts expanding forever.
Also shown is the average value of the scale factor, which is seen to be
potentially very different from the typical trajectories (and it is not even
the same as the trajectory having the same initial value of $a$, i.e., that
for which $a(t_0) = \langle a \rangle$).}
\label{Trajs}
\end{figure}

In most cases of cosmological relevance, even if one wants to compare
the canonical quantization procedure with less usual ones
\cite{Bergeron:2017ddo}, or even when a larger minisuperspace is
considered, e.g., to~account for a possible anisotropy (Bianchi Universe)
\cite{Peter:2016kan}, one ends up essentially solving a Schr\"odinger
equation in a potential. Figure \ref{Trajs} exemplifies a particular
case, showing the real and imaginary parts of the wave function, its
amplitude, and the derivative of its phase with respect to the scale
factor. It~illustrates that not only is the singularity resolved by
quantum mechanics effects in this approach, but that the resulting
actual trajectory depends crucially on the initial condition of the
scale factor.

\section{Conclusions}

The trajectory method, also known as de Broglie Bohm pilot wave, permits
a clearer understanding of how quantum effects may affect cosmology near
the singularity, resolving the latter. However, it also shows that
defining the state itself may not be sufficient, as the initial
condition fixes the subsequent evolution of the scale factor: it may
bounce once or many times depending on its initial value! If one
calculates perturbations in a self-consistent way \cite{Peter:2008qz} on
top of such a trajectory, they will depend explicitly on which
trajectory has been chosen. This could actually provide a means of
measuring the time evolution of the very early scale factor.
\vspace{6pt}

\acknowledgments{I gratefully acknowledge enlightening conversations with
J.-P. Gazeau, N. Pinto-Neto and particularly S. Vitenti, who also provided
figures. I would like to thank the Labex Institut Lagrange de Paris
(reference ANR-10-LABX-63) part of the Idex SUPER, within which this work
has been partly done. I am hosted at Churchill College, Cambridge, partially
supported by a fellowship funded by the Higher Education, Research and
Innovation Dpt of the French Embassy to the United-Kingdom.}

\conflictsofinterest{The author declares no conflict of interest.} 

\reftitle{References}

\end{document}